\documentclass[aoas]{imsart}

\RequirePackage{amsthm,amsmath,amsfonts,amssymb}
\RequirePackage[authoryear]{natbib}
\RequirePackage[colorlinks,citecolor=blue,urlcolor=blue]{hyperref}
\RequirePackage{graphicx}

\usepackage{dsfont}
\usepackage{amsmath,amssymb}
\usepackage{amsthm}
\usepackage[utf8]{inputenc}
\usepackage[T1]{fontenc}
\usepackage{hyperref}
\usepackage{url}
\usepackage{booktabs}
\usepackage{amsfonts}
\usepackage{nicefrac}
\usepackage{microtype}
\usepackage{graphicx}
\usepackage{xcolor}
\usepackage{lipsum}
\usepackage{float}
\usepackage{tabularx}
\usepackage{subcaption}
\usepackage{graphicx}
\usepackage{algorithm,algorithmic,amsmath}

\startlocaldefs
\theoremstyle{plain}

\theoremstyle{remark}

\newcommand{\bZ}{\mathbf{Z}}
\newcommand{\bY}{\mathbf{Y}}

\newcommand{\bp}{\mathbf{p}}
\newcommand{\bLambda}{\boldsymbol{\Lambda}}

\endlocaldefs

\begin{document}

\begin{frontmatter}
\title{An Empirical Bayes Approach for Constructing Confidence Intervals for Clonality and Entropy}
\runtitle{}

\begin{aug}
\author[A]{\fnms{}~\snm{Zhongren Chen}\ead[label=e1]{zhongren.chen@stanford.edu}},
\author[B]{\fnms{}~\snm{Lu Tian}\ead[label=e2]{lutian@stanford.edu}\orcid{0000-0000-0000-0000}}
\and
\author[C]{\fnms{}~\snm{Richard A. Olshen}\ead[label=e3]{olshen@stanford.edu}}
\address[A]{Department of Statistics,
Stanford University\printead[presep={,\ }]{e1}}

\address[B]{Department of Biomedical Data Science,
Stanford University\printead[presep={,\ }]{e2}}

\address[C]{Department of Biomedical Data Science,
Stanford University\printead[presep={,\ }]{e3}}
\end{aug}

\begin{abstract}
This paper is motivated by the need to quantify human immune responses to environmental challenges.  Specifically, the genome of the selected cell population from a blood sample is amplified by the well-known PCR process of successive heating and cooling, producing a large number of reads.  They number roughly 30,000 to 300,000 in our applications.  Each read corresponds to a particular rearrangement of so-called V(D)J sequences.   In the end, the observed data consist of a set of integers, representing numbers of reads corresponding to different V(D)J sequences.  The underlying relative frequencies of distinct V(D)J sequences can be summarized by a probability vector, with the cardinality being the number of distinct V(D)J rearrangements present in the blood. The statistical question is to make inferences on a summary parameter of the probability vector based on a single multinomial-type observation of large dimension. Popular summaries of the diversity of a cell population include clonality and entropy, or more generally, a suitable function of the probability vector. A point estimator of the clonality based on multiple replicates from the same blood sample has been proposed previously.  After obtaining a point estimator of a particular function, the remaining challenge is to construct a confidence interval of the parameter to appropriately reflect its uncertainty.  In this paper, we have proposed to couple the empirical Bayes method with a resampling-based calibration procedure to construct a robust confidence interval for different population diversity parameters. The method is illustrated via extensive numerical studies and real data examples.
\end{abstract}


\end{frontmatter}

\section{Introduction}
This paper is motivated by the need to construct confidence intervals for parameters summarizing the diversity of a cell population consisting of cells of different types.  We first introduce the sources of data and give some details here.  More biological and statistical background in the case of point estimation can be found in a previous paper \citep{tian2019clonality}. The problem has its biomedical origin in attempting to quantify human immune responses to a environmental challenge; more specifically, to quantify the adaptive immunologic response to any antigen, e.g., vaccination against Covid 19 virus.  
Briefly, blood is sampled from a patient.  This blood sample may be divided, as equally as possible, into several parts, i.e., replicates.  The genome of the selected cell subpopulation in each replicate is amplified by the well-known PCR process of successive heating and cooling.  One resultant from each replicate is then chosen without prejudice for sequencing, producing a large number of reads.  They number roughly 30,000 to 300,000 per replicate.  Each read corresponds to a particular rearrangement of the so-called V(D)J sequence. 
In the end, the observation from a particular replicate consists of a set of numbers of reads for different V(D)J sequences.  Mathematically, each observation can then be thought of as a finite dimensional random vector  $\bZ = (Z_1, Z_2, \cdots, Z_{C})'$ reflecting the underlying relative frequency of different cell subpopulations with particular V(D)J rearrangements in the entire circulation system from which the blood was sampled originally. According to the Rao-Blackwell theorem, the underlying relative frequencies can be summarized by a probability vector  $\bp = (p_1, p_2, \cdots, p_{C_0})',$ where $C_0,$ the cardinality of the vector $p,$ represents the total number of different V(D)J rearrangement in the circulation. While $C_0$ is finite, its value is unknown, since some of the V(D)J rearrangements may not be observed in the replicate due to their rarity. In other words, $C,$ the number of observed V(D)J rearrangements could be substantially smaller than the number of different V(D)J rearrangements in the blood. While we know from criticisms that the PCR process may favor some rearrangements over others, that distinction is ignored here. 

Statistical concerns are making inferences on a summary parameter based on this single multinomial-type observation, $\bZ$. One popular summary of the diversity of a cell population is its clonality, $G_C(\bp)=\|\bp\|_2^2 $, the squared $l_2$ norm of vector $\bp \in R^{C_0}$, which obviously varies between ${C_0}^{-1}$ and 1.   We have reported a study of point estimation for clonality based on multiple replicates for each blood sample in a previous paper \citep{tian2019clonality}.  After obtaining a point estimator of a particular function of $\bp$ such as clonality characterizing the composition of the cell population, the remaining challenge is to construct a confidence interval for the relevant parameter to appropriately reflect the uncertainty in this point estimator.  Consequently, in what follows, attention is devoted to interval estimation.  Specifically, we propose to study point and interval estimations for  clonality without requiring multiple replicates.  Another extremely popular measure for population diversity is entropy, which is defined as
 $$G_E(\bp)=\sum_{j=1}^{C_0} \left(-p_j \log p_j\right) $$ 
 (see, for example, \cite{glanville2017identifying}); the logarithm is taken to be to the natural base as the convention. Many authors prefer to summarize the variability of $\bp$ by its entropy.  Of course, the notion of entropy has a storied history in communications, indeed in many aspects of engineering \citep{zurek2018complexity}.  The estimation of entropy itself is very difficult \citep{chao2003nonparametric}.  Especially, we are unaware of careful attempts to form confidence intervals for entropy. The method proposed in this paper will cover the point and interval estimations for parameters such as entropy or other functions of the probability vector $\bp$.
 
We begin with an explanation of matters that bear upon parametric approaches in Section ~\ref{sec:Method}. All material applies to a fixed $\bp$. However, the key is that we assume that these $C_0$ components of the probability vector $\bp$ are realizations from a parametric ``prior'' distribution, which can be estimated based on observed data, although there is oftentimes no sufficient information in estimating the values of all individual components. With the estimated ``prior'' distribution generating the individual probability components, point and interval estimates for the function of interest can be obtained based on the ``posterior distribution'' derived from the Bayesian theorem. This is essentially an Empirical Bayesian approach, where the ``prior'' distribution generating the probability vector is empirically estimated.  However, such a naive interval may fail to cover the true value of the function with the desired coverage level due to the simple fact that the uncertainty of the estimated ``prior'' distribution is not considered in this Empirical Bayesian approach \citep{casella1985introduction, carlin2000empirical}. Therefore, we have proposed one additional calibration step to correct the under-coverage of this naive empirical Bayesian confidence interval.

We have realized that a good part of understanding the performance of the constructed confidence interval should be summarized by extensive computations on simulated clinical or other suitable data.  Some computations are given in the Numerical Study Section (Section~\ref{sec:Simulation Study}) and real data example Section (Section~\ref{sec: Real Data Studies}), that follow the Method Section (Section~\ref{sec:Method}).  
Additional suggestions for further research are topics of our Discussion Section(Section~\ref{sec:Discussion}).

\section{Method}\label{sec:Method}
\subsection{The General Framework}

The complete data consist of $n$ pairs of observations $\left\{(Z_i, Y_i), i=1,\cdots, n\right\}.$ Let $$(Z_i, Y_i)\overset{i.i.d.}{\sim} p(z, y \mid \theta_0), i=1, \cdots, n,$$ where $p(z, y\mid \theta_0)$ is the density function for the joint distribution with respect to $(Z_i, Y_i), i=1,\cdots, n.$ Suppose that we only observe $\bZ=\{Z_1, \cdots, Z_n\}$,  $\bY=\{Y_1, \cdots, Y_n\}$ is missing but correlated with $\bZ=\{Z_1, \cdots, Z_n\},$ and $\theta_0$ is a fixed but unknown parameter. Our aim is to construct a confidence interval covering $G(\bY)=G(Y_1, \cdots, Y_n)$ with a specified probability based on observed data $\bZ$ only,  where $G(\cdot)$ is a given function. Note that different from the conventional setting, where the parameter of interest is a deterministic population parameter,  $G(\bY)$ is a random variable that varies from data to data. For this reason, we use credible interval and confidence interval interchangeably in the current paper with a slight abuse of notations. 

Assuming that a point estimator $\hat{\theta}(\bZ)$ for $\theta_0$ based on $\bZ$ is available, we can derive the conditional distribution of $Y_i\mid Z_i, \theta_0=\hat{\theta}(\bZ), i=1, \cdots, n,$ and simulate multiple copies of $G^*=G(\bY)$ directly from the conditional distribution of  $G(\bY) \mid \bZ, \theta_0=\hat{\theta}(\bZ).$ A credible interval for $G(\bY)$ can then be constructed based on sampled $G^*$s.  See Algorithm~\ref{alg: Naive_EB} for this naive approach.

\begin{algorithm}
 \begin{algorithmic}[1]

\STATE Find a point estimator of $\theta_0,$ $\hat{\theta}(\bZ).$
\STATE Simulate $Y_i^*$ from the conditional distribution $Y_i|Z_i, \theta_0=\hat{\theta}(\bZ),$ for $i=1, \cdots, n.$
\STATE Compute $G^*=G(Y_1^*, \cdots, Y_n^*).$
\STATE Repeat steps 2 and 3 a large number of times, obtain $G^*_1, \cdots, G_B^*,$ and return the $2.5$th and $97.5$th percentiles of the empirical distribution of $\{G_1^*, \cdots, G^*_B\}$ as a $95\%$ credible interval for $G(\bY).$

\caption{A Naive Algorithm to Construct a $95\%$ credible Interval for $G(Y_1, Y_2, \cdots, Y_n)$}
\label{alg: Naive_EB}
\end{algorithmic}
\end{algorithm}
 
 In practice, however, the behavior of this interval is largely dependent on how well $\hat{\theta}(\bZ)$ is in estimating $\theta_0$. A significant discrepancy may be observed between $\theta_0$ and $\hat{\theta}(\bZ)$ resulting severe under-coverage.  To make the constructed intervals robust to the variability of $\hat{\theta}(\bZ)$, we propose to couple a method incorporating the variance of $\hat{\theta}(\bZ)$ and a calibration step using the parametric bootstrap method \citep{efron1994introduction}.
 
 Specifically, we first obtain an estimator of $\theta_0,$ $\hat{\theta}(\bZ),$ and an estimator of the variance of $\hat{\theta}(\bZ),$ $\hat{J}(\bZ).$  Then, we attempt to construct the credible interval based on a ``posterior'' distribution:
\begin{equation}
G(\bY)\mid \bZ, \theta_0=\theta^*;   \theta^* \sim N\left\{\hat{\theta}(\bZ), \hat{J}(\bZ)\right\}, \label{eq:posterior}
\end{equation} 
which, in general, has a larger variability than the naive ``posterior'' distribution
$$G(\bY)\mid \bZ, \theta_0=\hat{\theta}(\bZ)$$ 
and likely results in a wider credible interval. In addition, we also consider a calibration step based on parametric bootstrap \citep{efron1994introduction} similar to the correction introduced in \cite{carlin1990approaches}.  To be specific, we simulate ``observed'' data from an assumed model with $\theta_0=\hat{\theta}(\bZ)$,  simulate realizations of $G(\bY)$ from the ``posterior" distribution given above, and construct the $100(1-\alpha)$\% credible intervals for different $\alpha$ based on the quantiles of samples drawn from the posterior distribution. After repeating this simulation a large number of times, we will examine the empirical coverage level of the constructed credible interval with respect to the true $G(\bY)$ in the simulated data, anticipating that the real coverage level of $100(1-\alpha)$\% credible intervals may differ from $(1-\alpha).$  We will find the $\alpha$ level such that the corresponding $100(1-\alpha)$\% credible interval has a coverage level of 95\%. This $\alpha$ level will then be used to construct the credible interval based on the ``posterior'' distribution (\ref{eq:posterior}) from the original data. The mathematical rationale is that 
$$P\left\{G(\bY)< q_\alpha(\bZ) \mid \theta_0\right\}\approx P\left\{G(\bY)< q_\alpha(\bZ) \mid \hat{\theta}(\bZ)\right\},$$
where $q_\alpha(\bZ)$ is the $\alpha$th quantile of the ``posterior'' distribution (\ref{eq:posterior}), and the parametric bootstrap was used to estimate $P\left\{G(\bY)< q_\alpha(\bZ) \mid \hat{\theta}(\bZ)\right\}.$ This step can be viewed as a calibration step based on a bootstrap method \citep{efron1994introduction}. 
In the end, if denoting the resulting credible interval as $[\hat{L}(\bZ), \hat{U}(\bZ)],$ we expect that
$$ P\left\{\hat{L}(\bZ)\le G(\bY)\le \hat{U}(\bZ)\right\}=0.95,$$
where the probability is with respect to the joint distribution of $(\bY, \bZ).$ The detailed steps are outlined in algorithm~\ref{alg: General_EB}. 

\begin{algorithm}
 \begin{algorithmic}[1]

\STATE Obtain an point estimator $\hat{\theta}(\bZ)$ of $\theta_0.$
\STATE Compute a consistent variance estimator of $\hat{\theta}(\bZ), $  $\hat{J}(\bZ).$
\FOR{$b=1, 2, \cdots, B,$}
\STATE Simulate $\tilde{\theta}_{1(b)},...,\tilde{\theta}_{n(b)}\overset{i.i.d.}{\sim}N(\hat{\theta}(\bZ), \hat{J}(\bZ)).$
\STATE Simulate $Y_{1(b)}^* \sim p_Y\left(y \mid Z_{1}, \tilde{\theta}_{1(b)}\right), \cdots, Y_{n(b)}^* \sim p_Y\left(y \mid Z_{n}, \tilde{\theta}_{n(b)}\right),$ where $p_Y(y\mid z, \theta)$ is the density function of $Y$ conditional on $Z=z$ and $\theta_{0}=\theta.$
\STATE Compute $G_b^*(\bZ)=G\left(Y_{1(b)}^*, \cdots, Y_{n(b)}^*\right).$
\ENDFOR
\FOR{$i=1, 2, \cdots, R,$}
\STATE Simulate a new dataset $\{Y_{ij}^{\ast}, Z_{ij}^{\ast}\}_{j=1}^n \overset{i.i.d.}{\sim}p\left(y,z \mid \hat{\theta}(\bZ) \right)$. Let $\bZ_{i}^{\ast}=\{ Z_{ij}^{\ast}\}_{j=1}^n$ and $\bY_{i}^{\ast}=\{Y_{ij}^{\ast}\}_{j=1}^n.$
\STATE Calculate the random variable to be estimated: $G_i^*=G(Y_{i1}^*, \cdots, Y_{iB}^*)=G(\bY_i^*).$
\STATE Obtain the point estimator $\hat{\theta}(\bZ^*_i)$ of $\hat{\theta}$ based on the generated data $\bZ^*_i.$
\STATE Obtain a variance estimator of $\hat{\theta}(\bZ^*_i)$, $\hat{J}(\bZ^*_i).$
\STATE Using Steps 3-6 to obtain $\left\{G_1(\bZ^*_i), \cdots, G_B(\bZ^*_i)\right\}.$
\STATE Construct the $100(1-\alpha)\%$ confidence interval as the $100\alpha/2$th percentile and $100(1-\alpha/2)$th percentile of $\left\{G_1(\bZ^*_i), \cdots, G_B(\bZ^*_i)\right\}$, denoted by $\widehat{CI}_i(1-\alpha).$
\ENDFOR
\STATE Calculate the empirical coverage level of $\widehat{CI}_i(1-\alpha)$ as
$$ \frac{1}{R}\sum_{i=1}^R I\left(G_i^* \in \widehat{CI}_i(1-\alpha)\right),$$
where $I(\cdot)$ is an indicator function.
\STATE Determine the value of $\alpha_0$ such that the empirical coverage level of $\widehat{CI}_i(1-\alpha_0)$ is $95\%,$ i.e., 
$$ \frac{1}{R}\sum_{i=1}^R I\left(G_i^* \in \widehat{CI}_i(1-\alpha_0)\right)=0.95$$
\STATE Return the 95\% credible interval for $G(\bY)$ based on observed data as the interval between the $100\alpha_0/2$th and $100(1-\alpha_0/2)$th quantiles of $\left\{G_1^*(\bZ), \cdots, G_B^*(\bZ)\right\}.$
\caption{A General Algorithm to Construct a $95\%$ Confidence Interval for $G(\lambda)$}
\label{alg: General_EB}
\end{algorithmic}
\end{algorithm}


\subsection{Interval Estimates of Clonality and Entropy}
 We now apply the general algorithm introduced in the previous section to constructing confidence intervals for clonality and entropy. Consider a parametric model for the number of cells from a clone, i.e., cells with a particular V(D)J rearrangement:
 $$
\lambda_i \sim Gamma(a_0, b_0),\text{ }i=1, ..., C_0,
$$
$$
Z_i \sim Poisson(\lambda_i),\text{ }i = 1, ... , C_0,
$$
where $C_0$ is the total number of clones. $\{\lambda_i\}_{i=1}^{C_0}$ and $(a_0, b_0)$ are unknown to us, but we observe $Z_i$ if $Z_i>0$, which is the number of reads corresponding to the $i$th V(D)J rearrangement.  Therefore, the observed data consist of the truncated independent Poissons with parameters $\lambda_i, ...,\lambda_{C_0}$:
$$
\left\{Z_i|Z_i>0,\text{ }1\le i\le C_0\right\}
$$

Here $\left\{Z_i|Z_i=0,\text{ }1\le i\le C_0\right\}$ are missing data. Now, we define clonality as $G_C$ and entropy as $G_E:$ 
$$
G_C:=\sum_{i=1}^{C_0}{\biggl(\frac{{\lambda_i}}{\sum_{j=1}^{C_0}\lambda_j}\biggl)}^2,
$$ and 
$$
G_E:=-\sum_{i=1}^{C_0}\biggl\{{\frac{{\lambda_i}}{\sum_{j=1}^{C_0}\lambda_j}}\log\biggl({\frac{{\lambda_i}}{\sum_{j=1}^{C_0}\lambda_j}}\biggl)\biggl\}.
$$
Both are functions of $\left\{\lambda_1, \cdots, \lambda_{C_0}\right\}.$
We aim to obtain point estimations as well as confidence intervals for $G_C$ and $G_E$.

To fit into the general framework described above, we use an estimator of $C_0,$ denoted by $\hat{C},$ to replace $C_0$. Then  $\bZ$ corresponds to data $\left\{Z_1, \cdots, Z_{\hat{C}}\right\}$, which are partially observed,  $\bY$ corresponds to $\left\{\lambda_1, \cdots, \lambda_{\hat{C}}\right\}$ and $\theta_0$ corresponds to $(a_0, b_0)$, the shape and rate parameters in the gamma distribution. Below, we outline the algorithm step by step.

\subsubsection{Estimation of $(a_0, b_0)$ and $C_0$}
Marginally, $Z_i$ follows a negative binomial distribution, i.e., 

\begin{align*} 
\mathbb{P}(Z_i=z) = p_Z(z|a_0, b_0)&= \int_0^{\infty}\frac{e^{-\lambda}{\lambda}^z}{z!}\frac{\lambda^{a_0-1}e^{-b_0\lambda}b_0^{a_0}}{\Gamma(a_0)}d\lambda \\ 
&= {z+a_0-1 \choose z}\biggl(\frac{b_0}{b_0+1}\biggl)^{a_0}\biggl(\frac{1}{b_0+1}\biggl)^z,
\end{align*}
where $p_Z(\cdot \mid a, b)$ represents the density function of the distribution of $Z_i.$
We aim to first estimate the parameters $a_0$ and $b_0$ in the gamma distribution based on observed data consisting of positive reads 
$$\bZ_{O}=\left\{Z_i\mid Z_i>0, i=1,\cdots, C_0\right\}.$$  
The resulting log-likelihood function in terms of $a_0$, $b_0$ is

\begin{equation} \label{eqn: marginal likelihood}
l(a, b)=\sum_{z_i\in \bZ_{O}}\log\biggl\{{\frac{p_Z(z_i\mid a,b)}{1-p_Z(0 \mid a,b)}}\biggl\},
\end{equation}
which can be maximized via an EM algorithm \citep{dempster1977maximum, mclachlan1988fitting}. 

If $C_0$ is known, i.e., we observed clones with $Z_i=0$, the full log-likelihood function becomes

\begin{equation} \label{eqn: full likelihood}
\sum_{z_i \in \bZ_O}\log\{{p(z_i|a_0,b_0)}\}+\sum_{Z_i=0}\log\{{p(0|a_0,b_0)}\}.
\end{equation}
Therefore, the EM algorithm consists of the following E and M steps.

\begin{itemize}
\item E-step: For a given estimator $\hat{a}$ and $\hat{b}$, estimate
$$
\mathbb{E}[C_0-C\mid\hat{a}, \hat{b}]
$$
as
$$
\mathbb{E}[C_0-C\mid\hat{a}, \hat{b}]= \frac{p_Z(0\mid\hat{a}, \hat{b})}{1-p_Z(0\mid\hat{a}, \hat{b})}C
$$
based on the relationship
$$
p_Z(0\mid\hat{a}, \hat{b})(\mathbb{E}[C_0-C\mid\hat{a}, \hat{b}]+C) = \mathbb{E}[C_0-C\mid\hat{a}, \hat{b}] , 
$$ 
where $C$ is the number of observed nonzero $Z_{i}s$. Since
$$
p_Z(0\mid a_0, b_0) = \Bigl(\frac{b_0}{b_0+1}\Bigl)^{a_0},
$$
we have 
$$
\hat{n}_0 = \mathbb{E}[C_0-C\mid\hat{a}, \hat{b}]=\frac{\hat{b}^{\hat{a}}C}{{(\hat{b}+1)^{\hat{a}} - \hat{b}^{\hat{a}}}}
$$

\item M-step: Update the MLE of $a_0$, $b_0$ by maximizing
$$
\sum_{z_i\in \bZ_O}\log\{{p_Z(z_i\mid a_0,b_0)}\}+\hat{n}_0\log\{p_Z(0\mid a_0,b_0)\}.
$$
This optimization can be achieved via an \textbf{Inner EM algorithm} treating $\{\lambda_i\}$ as missing variables:
\begin{enumerate}
\item E-step: For given ($\hat{a}$, $\hat{b}$)
\begin{align*}
\mathbb{E}[\lambda_i\mid\hat{a}, \hat{b}]&=\frac{Z_i+\hat{a}}{1+\hat{b}}, \\
\mathbb{E}[\log(\lambda_i)\mid\hat{a}, \hat{b}]&= \Psi(Z_i+\hat{a})-\log(1+\hat{b}),\text{ }i = 1, ..., C,\\
\mathbb{E}[\lambda_0\mid\hat{a}, \hat{b}]&= \frac{\hat{a}}{1 + \hat{b}},\\
\mathbb{E}[\log(\lambda_0)\mid\hat{a}, \hat{b}]&= \Psi(\hat{a})-\log(1+\hat{b}), 
\end{align*}
where $\lambda_0$ corresponds to the Poisson rate of unobserved clones with $Z_i = 0$ and $\Psi(\cdot)$ is the dgamma function, the derivative of the log-transformed Gamma function, i.e., 
$\Psi(x) = \left[\log\left\{\int_0^\infty t^{x-1}e^{-t}dt\right\}\right]'.$
\item M-step: Maximize the log-likelihood function
\begin{align*}
&l(a_0, b_0) \\
&= (a_0 - 1)\Bigr[\sum_{i=1}^{C}\mathbb{E}[\log(\lambda_i)\mid\hat{a}, \hat{b}] + \hat{n}_0\mathbb{E}[\log(\lambda_0)\mid\hat{a}, \hat{b}]\Bigr] \\
& -(C + \hat{n}_0)[\log\Gamma(a_0) - a_0\log(b_0)] \\
& - b_0\Bigr[\sum_{i=1}^{C}\mathbb{E}[\lambda_i\mid\hat{a}, \hat{b}] + \hat{n}_0\mathbb{E}[\lambda_0\mid\hat{a}, \hat{b}]\Bigr] \\
&=(a_0-1)\Bigr[\sum_{i=1}^C\bigr\{\Psi(z_i + \hat{a}) - \log(1 + \hat{b})\bigr\} + \hat{n}_0\bigr\{\Psi(\hat{a}) - \log(1 + \hat{b})\bigr\}\Bigr] \\
&- (C + \hat{n}_0)[\log\Gamma(a_0) - a_0\log(b_0)] - b_0\frac{\sum_{i=1}^{C}(z_i + \hat{a}) + \hat{n}_0\hat{a}}{1 + \hat{b}}
\end{align*}
\end{enumerate}
\end{itemize}

In practice, we iterate the inner EM algorithm to maximize
$$
\sum_{z_i\in \bZ_O}\log\{{p_Z(z_i\mid a_0,b_0)}\}+\hat{n}_0\log\{p_Z(0\mid a_0,b_0)\}
$$
for a given $\hat{n}_0$ and iterate the outer EM algorithm until $\hat{n}_0$ converges. The final convergence of $(\hat{a}, \hat{b}, \hat{n}_0)$ can be assessed by the relative change of the log-likelihood:

$$
\sum_{z_i \in bZ_O}\log\biggl\{{\frac{p_Z(z_i\mid\hat{a}, \hat{b})}{1-p_Z(0\mid\hat{a}, \hat{b})}}\biggl\}.
$$
whose value should increase with each iteration of the outer EM algorithm.


\subsubsection{Estimation of the variance of $(\hat{a}, \hat{b})$ }
After obtaining the MLE $(\hat{a},\hat{b})$ for $(a_0,b_0)$ using the EM algorithm, 
we can estimate the variance of $(\hat{a}, \hat{b})$ by the inverse of the observed information matrix, which can be calculated by taking the second derivative of the log-likelihood function w.r.t. $a$ and $b,$ i.e.,  
$$
\hat{J}(\bZ_O)= 
\begin{pmatrix}
 \frac{\partial^2l(a,b)}{\partial a^2}  & \frac{\partial^2l(a,b)}{\partial a \partial b}  \\
\frac{\partial^2l(a,b)}{\partial b \partial a}  &  \frac{\partial^2l(a,b)}{\partial b^2} 
\end{pmatrix}\Bigr\rvert_{(a,b)=(\hat{a},\hat{b})}^{-1},
$$
where $l(a, b)$ is given in (\ref{eqn: marginal likelihood}).

\subsubsection{Sampling from the posterior distribution}
In constructing the 95\% confidence interval, we need to generate 
$$
\left(\begin{array}{c}a^*_i\\ b^*_i \end{array}\right)\sim \exp\Biggl\{N\Biggl(\begin{pmatrix}
{\hat{a}} \\
{\hat{b}}  
\end{pmatrix}, 
\hat{A}(\bZ_O)
\hat{J}(\bZ_O)
\hat{A}(\bZ_O)\Biggl)\Biggl\}
$$
and
$$\lambda_i^* \sim \lambda_i\mid Z_i, a_i^*, b^*_i; $$
which is a gamma distribution with the shape and rate parameters being 
$a_i^*+Z_i$ and $b^*_i+1,$ respectively, where $\hat{A}(\bZ_O)=\mbox{diag}(\hat{a}^{-1}, \hat{b}^{-1}).$ This approach ensures that sampled $a^*_i$ and $b^*_i$ are always positive. Operationally, we pretend that $C_0=\hat{C}$ and let 
$$\bZ=\left\{Z_1, Z_2, \cdots, Z_C, Z_{C+1}=0, \cdots, Z_{\hat{C}}=0\right\}=\bZ_O\cup\{0, \cdots, 0\}.$$
and 
$$\bY=\left\{\lambda_1, \lambda_2, \cdots, \lambda_{\hat{C}}\right\}.$$
The credible intervals $\widehat{CI}_C(\bZ_O)$ and $\widehat{CI}_E(\bZ_O)$ are calibrated such that 
$$ P\left\{\sum_{i=1}^{\hat{C}} \left(\frac{\lambda_i}{\sum_{j=1}^{\hat{C}}\lambda_j} \right)^2 \in \widehat{CI}_C(\bZ_O) \biggm | \hat{a}, \hat{b}\right\}=0.95$$
and
$$ P\left\{\sum_{i=1}^{\hat{C}} -\frac{\lambda_i}{\sum_{j=1}^{\hat{C}}\lambda_j}\log\left(\frac{\lambda_i}{\sum_{j=1}^{\hat{C}}\lambda_j} \right)^2 \in \widehat{CI}_E(\bZ_O)\biggm| \hat{a}, \hat{b}\right\}=0.95$$


The full algorithm is provided in Algorithm \ref{alg: main_algorithm}.
Note the algorithm for constructing confidence intervals for entropy is the same as that for clonality except that we replace all $G_C(\bY)=\sum_{i=1}^{C_0}{\biggl(\frac{{\lambda_i}}{\sum_{j=1}^{C_0}\lambda_j}\biggl)}^2$ by $G_E(\bY) =-\sum_{i=1}^{C_0}\left\{{\frac{{\lambda_i}}{\sum_{j=1}^{C_0}\lambda_j}}\log\biggl({\frac{{\lambda_i}}{\sum_{j=1}^{C_0}\lambda_j}}\biggl)\right\}$.

\begin{algorithm}
 \begin{algorithmic}[1]
\STATE Use the proposed EM algorithm to obtain the MLE for $a_0$, $b_0$, and $C_0$, denoted by $\hat{a}$, $\hat{b}$ and $\hat{C}$, respectively.
\STATE Compute $\hat{J}(\bZ_O)$ as the inverse of the observed information matrix for $a_0$ and $b_0$ and $\hat{A}(\bZ_O)$
\STATE Let $\bZ=\left\{Z_1, \cdots, Z_{C}, Z_{C+1}=0, \cdots, Z_{\hat{C}}=0\right\}.$ 
\FOR{$b=1, \cdots, B,$}
\STATE Simulate $\left(\begin{array}{c}a_{i(b)}^*\\ b_{i(b)}^*\end{array}\right)\overset{i.i.d.}{\sim} \exp\Biggl\{N\Biggl(\begin{pmatrix}
{\hat{a}} \\
{\hat{b}}  
\end{pmatrix}, 
\hat{A}(\bZ_O)\hat{J}(\bZ_O)\hat{A}(\bZ_O)
\Biggl)\Biggl\}$.
for $i=1, \cdots, \hat{C},$
\STATE Simulate $\lambda_{i(b)}^*\sim Gamma\left(a^*_{i(b)}+z_{i}, b^*_{i(b)}+1\right)$ for $i = 1,...,\hat{C}$. 

\STATE Compute $G_{C(b)}^*(\bZ_O) =\sum_{i=1}^{\hat{C}}{\biggl(\frac{{\lambda_{i(b)}^*}}{\sum_{j=1}^{\hat{C}}\lambda_{i(b)}^*}\biggl)}^2
$.
\ENDFOR
\FOR{$i=1,\cdots, R,$}
\STATE Simulate $(\lambda_{i1}^{\ast},...\lambda_{i\hat{C}}^{\ast})\sim Gamma(\hat{a}, \hat{b})$ and let $\bLambda_i^*:=\left\{\lambda_{ij}^* \mid j=1, \cdots, \hat{C}\right\}$. 
\STATE Compute $G_{C}(\bLambda_i^*)=\sum_{j=1}^{\hat{C}}{\biggl(\frac{{\lambda_{ij}^*}}{\sum_{k=1}^{\hat{C}}\lambda_{ik}^*}\biggl)}^2
$.
\STATE Simulate a new set of data $\bZ^*_i=\{Z_{ij}^* \mid j=1, \cdots, \hat{C}\}$ where $Z_{ij}^*\sim Poisson(\lambda_{ij}^*)$. 
\STATE Induce the observed data $\bZ^*_{iO}=\{Z_{ij}^* \mid Z_{ij}>0, j=1, \cdots, \hat{C}\}$.

\STATE Repeat steps 4-5 for $\bZ^*_{iO}$ and obtain $\left\{G^*_{C(1)}\left(\bZ_{iO}^*\right), \cdots, G^*_{C(B)}\left(\bZ_{iO}^*\right)\right\}.$
\STATE Let $\widehat{CI}_{\alpha}(\bZ_{iO}^*)$ be the interval between the $100\alpha/2$th and $100(1-\alpha/2)$th percentiles of $\left\{G^*_{C(1)}\left(\bZ_{iO}^*\right), \cdots, G^*_{C(B)}\left(\bZ_{iO}^*\right)\right\}$ for various $\alpha.$ 
\ENDFOR
\STATE Determine the value of $\hat{\alpha}_0$ such that the proportion of $G_C(\bLambda_i^*)$ falls in  $\widehat{CI}_{\hat{\alpha}_0}(\bZ_{iO}^*)$  is closest to $95\%,$ i.e., 
$$ \frac{1}{R}\sum_{i=1}^n I\left\{G_C(\bLambda_i^*) \in \widehat{CI}_{\hat{\alpha}_0}(\bZ_{iO}^*)\right\}\approx 0.95.$$
\STATE Return the $100\alpha/2$th and $100(1-\alpha/2)$th percentiles of 
$\left\{G_{C(b)}^*(\bZ_O), b=1, \cdots, B\right\}$, which is  our $95\%$ confidence interval for the clonality.

\caption{The Algorithm to Construct a 95\% Confidence Interval for Clonality}
\label{alg: main_algorithm}
\end{algorithmic}
\end{algorithm}


\section{Simulation Study}\label{sec:Simulation Study}

 We have conducted a comprehensive simulation study to examine the empirical performance of the constructed confidence intervals for both clonality and entropy.  For each given set of $(a_0, b_0, C_0)$, we have repeated the experiments 500 times to compute the empirical coverage level of the constructed confidence intervals. In constructing the confidence interval, we set the number of resampling for the calibration to be $R=200.$ The number of posterior sampling was $B=500.$ $C_0$, the number of clones was set at $10,000.$ A higher $C_0$ is expected to yield better confidence intervals. The values of $a_0$ and $b_0$ were fixed at their maximum likelihood estimators based on data examples in Section~\ref{sec: Real Data Studies} to mimic real practice.  For comparison purpose, we have also constructed the confidence interval based on the naive Bayesian procedure given in Algorithm \ref{alg: Naive_EB} and the confidence interval based on the posterior distribution \ref{eq:posterior} without the calibration step.  
 
\begin{figure}
\includegraphics[width=5 in ]{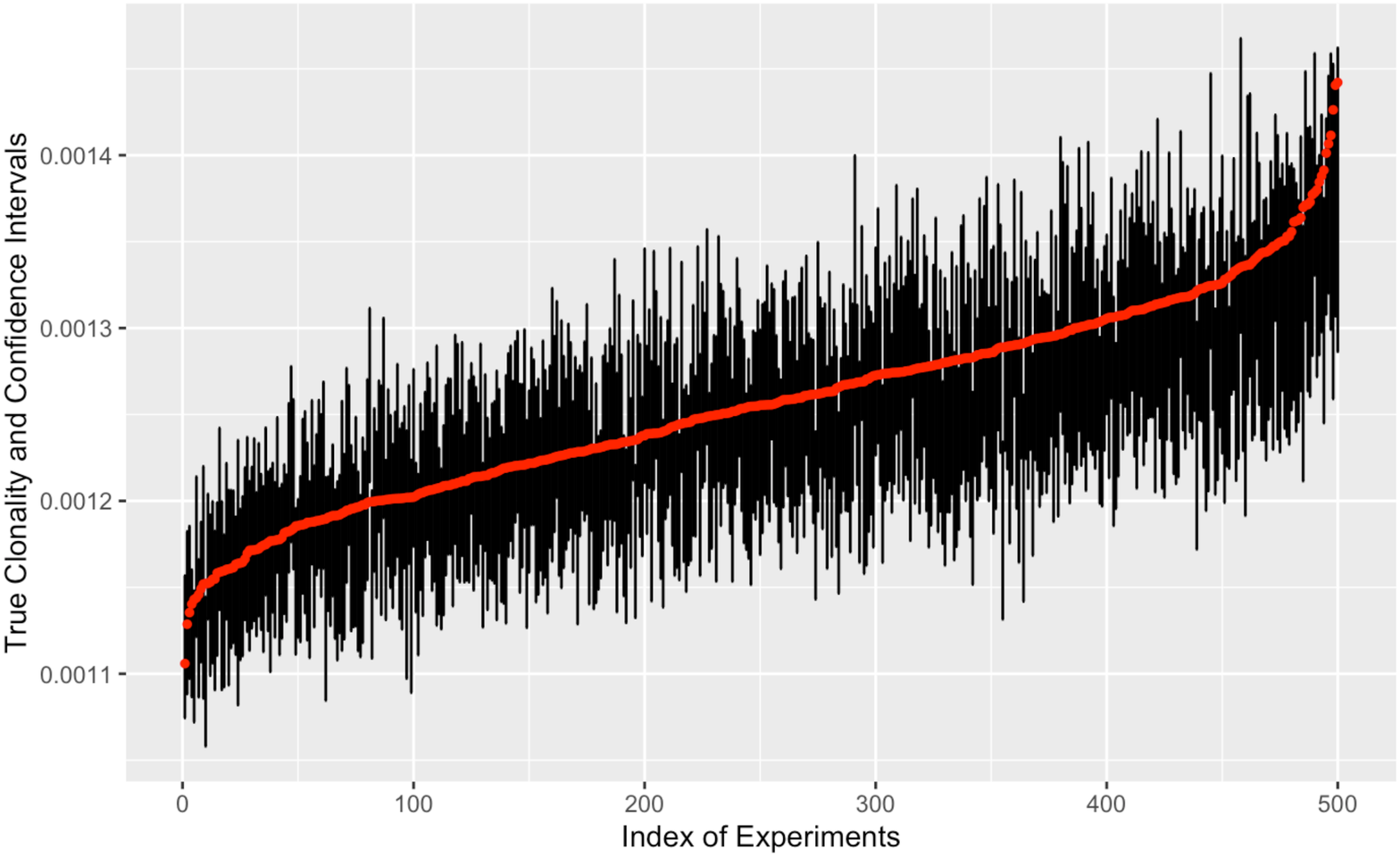}
\caption{95\% confidence intervals for the clonality based on 500 simulated datasets; The intervals are sorted based on the size of their true clonalities (from the smallest to the largest)}
\label{fig:simuci}
\end{figure}

 The simulation results were summarized in Table~\ref{tab:simulation}. The empirical coverage level of our proposed confidence intervals was close to its nominal level.  On the other hand, the confidence intervals based on the Naive EB approach undercovered the true parameter, which was expected considering the fact that the variability of $(\hat{a}, \hat{b})$ was ignored in constructing the confidence interval.  The coverage level of confidence intervals based on the posterior distribution \ref{eq:posterior} directly was higher than the nominal level, suggesting that those intervals were too conservative. It could be due to the fact that this method did not consider the fact that $(\hat{a}, \hat{b})$ was also depended on $\bZ_O$ and thus was correlated with quantiles of the posterior distribution.  This observation confirmed the essential role played by the calibration step, which was also the most computationally intensive part of the proposed algorithm.  We have also plotted confidence intervals for the clonality and their corresponding true clonalities from 500 data simulated with $(a_0, b_0, C_0)= (0.086, 0.111, 10,000)$ in Figure \ref{fig:simuci}. The confidence intervals are sorted according to the size of true clonalities.  It is clear that, unlike the conventional case, the true clonalities have quite substantial variation relative to the width of the 95\% confidence intervals, highlighting the importance of treating $G(\bY)$ as a random quantity.  

\begin{table}
  \centering
  \caption{The simulation results on empirical coverage level of constructed confidence intervals }\label{tab:simulation}
  \begin{tabular}{ccccl}
    \toprule
    $(a_0, b_0)$ &Method &Clonality Coverage& Entropy Coverage \\

    \midrule
                     &EB w Calibration             &94.0\% & 92.2\%                    \\
    $(0.732, 0.882)$ &EB w/o Calibration &100\%  & 100\%            \\
                     &Naive EB         &63.2\% & 45.6\%                 \\
    \midrule
                     &EB w Calibration              &91.2\% &95.8\%                    \\
    $(0.414, 0.335)$ &EB w/o Calibration &100\%  &100\%           \\
                     &Naive EB         &78.0\% &69.4\%                  \\

    \midrule
                     &EB w Calibration              &98.4\% & 96.6\%\\
    $(0.596, 0.960)$ &EB w/o Calibration &100\%  &100\%              \\
                     &Naive EB         &67.2\% & 40.6\%                \\
   \midrule
                     &EB w Calibration              &97.0\%  &96.0\%                    \\
    $(0.551, 0.775)$ &EB w/o Calibration &100\%   &100\%              \\
                     &Naive EB         &68.4\%  &44.8\%                \\
    \midrule
                     &EB w Calibration             &98.4\%      &99.2\%                    \\
    $(0.171, 0.301)$ &EB w/o Calibration &100\%       &100\%              \\
                     &Naive EB         &88.6\%      &70.6\%               \\
    \midrule
                     &EB w Calibration              &95.8\%      &98.2\%                    \\
    $(0.126, 0.132)$ &EB w/o Calibration &100\%      &100\%              \\
                     &Naive EB         &94.0\%      &83.8\%                \\
    \midrule
                     &EB  w Calibration             &94.4\%      &99.6\%                    \\
    $(0.0860, 0.111)$ &EB w/o Calibration &100\%       &100\%              \\
                     &Naive EB        &94.2\%      & 84.8\%                \\
    \midrule
                     &EB w Calibration              &95.6\%      &98.6\%                    \\
    $(0.113, 0.142)$ &EB w/o Calibration &100\%       &100\%              \\
                     &Naive EB         &91.2\%      &83.2\%                \\
  \bottomrule
\end{tabular}
\end{table}

 Next, we examined the empirical performance of the proposed confidence intervals when the underlying rates $\lambda_i, i=1, \cdots, C_0$ do not follow a gamma distribution, i.e., the assumed Poisson-Gamma model was misspecified in characterizing the data generation process. Specifically, we simulated $\lambda_i$ from a log-normal distribution $\exp\{N(\mu_0, \sigma_0^2)\}$ with chosen $\mu_0$ and $\sigma_0$ and generated the observed reads $\bZ_O.$ Using the same steps described above, we constructed confidence intervals based on 500 simulated datasets and calculated the empirical coverage level of these confidence intervals. The results are summarized in Table~\ref{tab:misspecification}.  The empirical coverage level of entropy was quite close to 95\% even though the Poisson-Gamma parametric model was misspecified, implying the robustness of the proposed confidence intervals. On the other hand, the constructed confidence interval for clonality severely underestimate the true parameter suggesting the importance of assuming an appropriate distribution of the Poisson rates. 

\begin{table}
  \centering
  \caption{The simulation results when the underlying model for rate is log-normal rather than Gamma}\label{tab:misspecification}
  \begin{tabular}{ccccl} 
    \toprule
    &$(\mu_0, \sigma_0^2)$ &Coverage for Clonality& Coverage for Entropy \\
    \midrule
    &$(-1.38, 1.64^2)$ &80.4\% & 98.2\% \\
    \midrule
    &$(-1.27, 1.72^2)$ &86.8\% & 97.8\% \\
    \midrule
    &$(-1.22, 1.50^2)$ &80.2\% &  98.4\%\\
    \midrule
    &$(-1.02, 1.62^2)$ &82.7\% &  95.8\%\\
  \bottomrule
\end{tabular}
\end{table}

  



\section{Real Data Studies}\label{sec: Real Data Studies}

\begin{figure}
\includegraphics[width=5 in ]{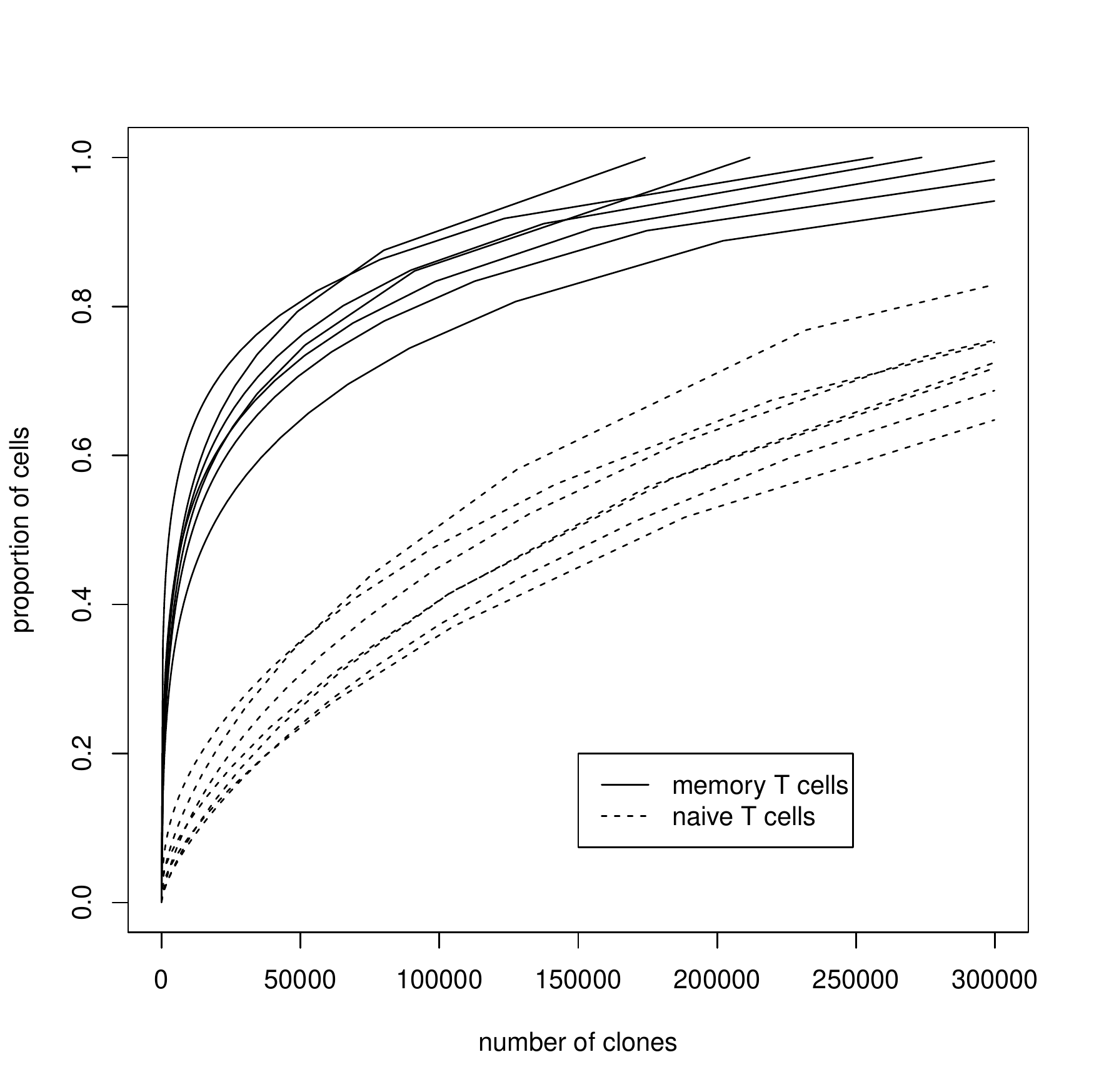}
\caption{Cumulative proportion of cells from clones sorted by the clone size from the biggest to the smallest}
\label{fig:dist}
\end{figure}

In this section, we illustrate our approach by applying it to analyze a
recent study conducted by \cite{qi2014diversity}. The objective was to investigate human T cell receptor (TCR) diversity. Specifically, we were interested in measuring TCR diversity by clonality and entropy.  In the study conducted by \cite{qi2014diversity}, five replicate TCR libraries of CD4 naive T cells and CD4 memory T cells are sequenced from each of seven participants. The total number of reads varied from $8.9 \times 10^4$ to $7.4 \times 10^5.$ First, we counted the total number of reads in each clone across five replicates, and Figure \ref{fig:dist} shows the observed cumulative proportions of clones sorted from the largest to the smallest for both CD4 naive and memory T cells. From the Figure, it is clear that large clones contain a bigger proportion of CD4 memory T cells than naive CD4 T cells, reflecting the relative evenness of the distribution of clone sizes of naive T cells. 

\begin{figure}
\includegraphics[width=5 in ]{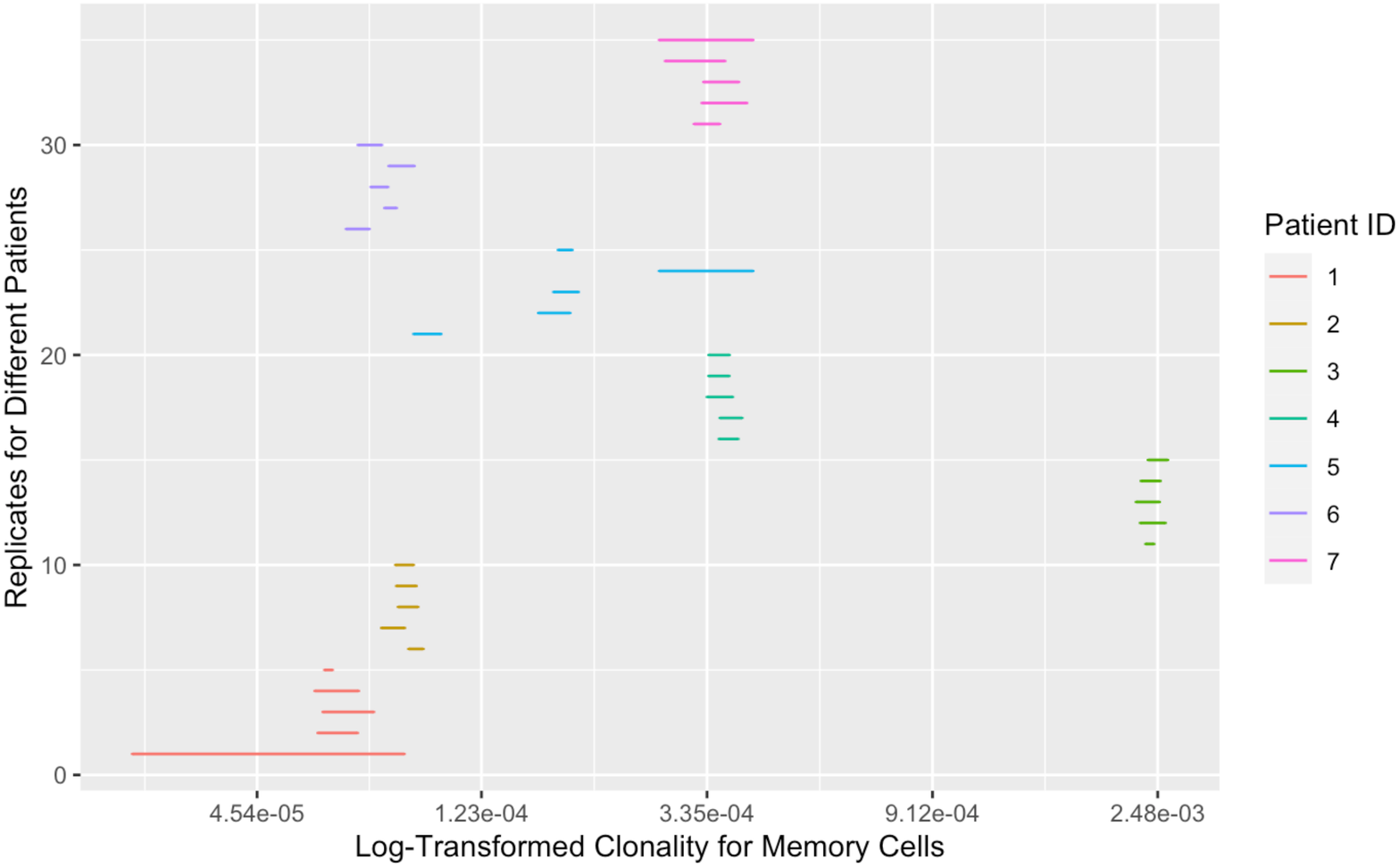}
\caption{95\% confidence intervals for clonality based on five replicates per patient (in the log scale for better visualization)}
\label{fig:ci1}
\end{figure}

\begin{figure}
\includegraphics[width=5 in ]{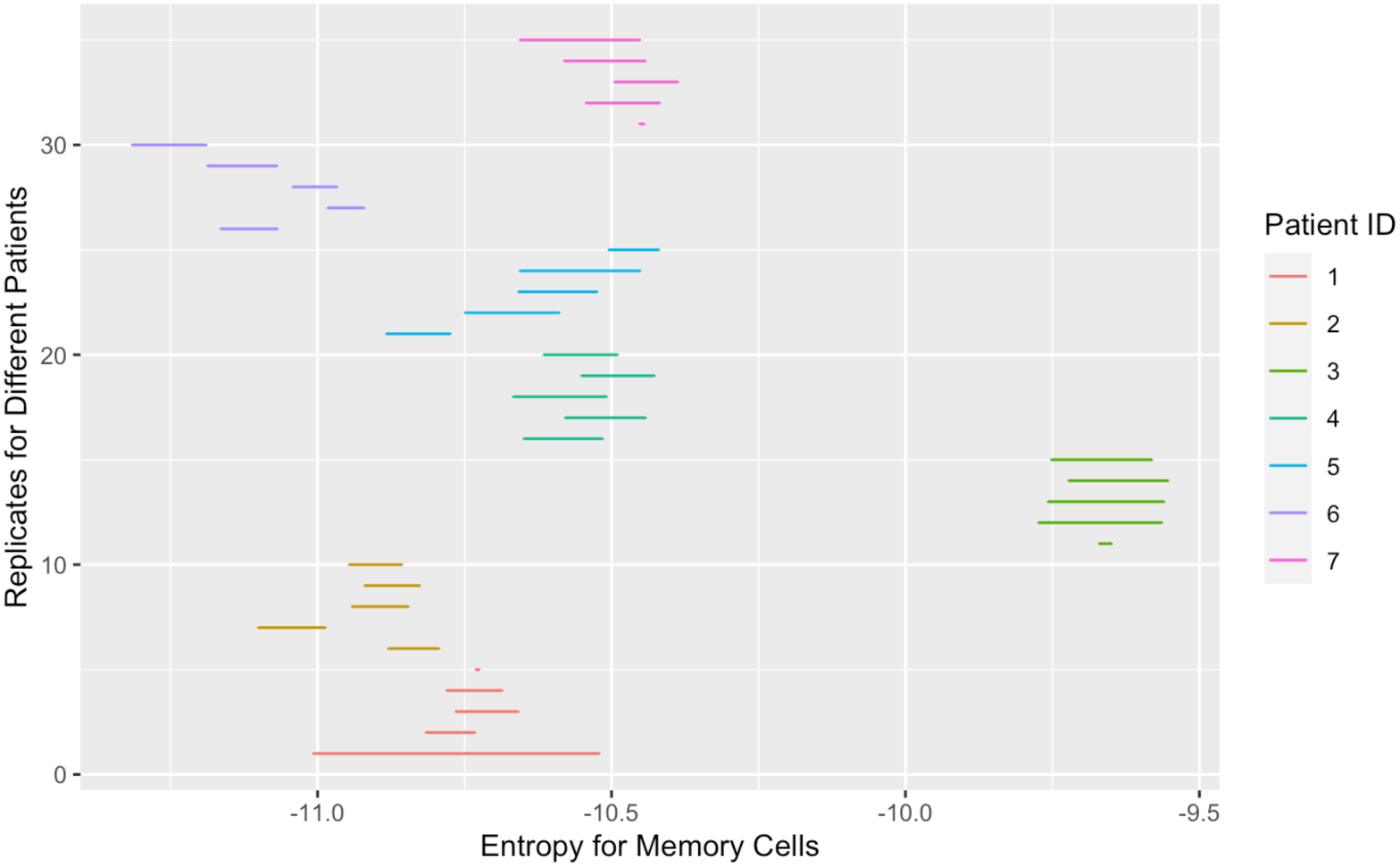}
\caption{95\% confidence intervals for Entropy based on five replicates per patient}
\label{fig:ci2}
\end{figure}

Next, we apply the proposed method to construct confidence intervals of log-transformed clonality and entropy based on data from five replicates per patient, which are plotted in Figures \ref{fig:ci1} and \ref{fig:ci2}.  It is clear that other than a few outliers the confidence intervals based on different replicates of the same participant are fairly consistent, suggesting a low within-person variation relative to between-person variation supporting the validity of the experiment result. The clonality of CD4 memory T cells is substantially greater than that of the CD4 naive T cells, again confirming observations on the evenness of the distribution of the naive T cell clones in Figure \ref{fig:dist}.  

Lastly, we estimate the ``average'' (log-transformed) clonality and entropy of naive CD4 T cells for three participants younger than 40 and four participants older than 70, separately,  based on a random effects model used in meta analysis. We have then compared the clonality and entropy between younger and older participants.  The average log-transformed clonality for memory T cells was -9.45[-9.66 to -9.24] for young participants and -7.79[-9.09 to -6.50] for old participants with a two-sided p-value of 0.014 for two group comparison. The average entropy is -10.97[-11.17 to -10.76] for young participants and -10.37 [-10.87 to 9.88] for old participants with a two-sided p-value of 0.029 for two group comparison.  These results suggest that immune diversity in old participants is lower than that in young participants, as anticipated, implying the aging effect on human immune system.


\begin{table}
  \centering
  {\small
  \caption{Inference for clonality and entropy of naive T cells}\label{tab:naive}
\begin{tabular}{ccccll}
    \toprule
    &$\hat{a}$ &$\hat{b}$& $\hat{C}$ &Clonality(95\% CI)$^*$  &Entropy(95\% CI)\\

    \midrule
     Patient 1 &0.732 &0.882 &446805 &5.40(5.27, 5.55) &-12.46(-12.48, -12.44)\\
    \midrule
     Patient 2 &0.414 &0.335 &441682 &8.43(8.31, 8.55) &-12.14 (-12.15, -12.13) \\ 
    \midrule
     Patient 3 &0.596 &0.960 &506737 &6.57(5.99, 7.20) &-12.47(-12.54, -12.41)\\
    \midrule
     Patient 4 &0.551 &0.775 &406797 &8.90(8.45, 9.33) &-12.21(-12.24, -12.18)\\

    \midrule
     Patient 5 &0.347 &0.473 &516751 &11.6(11.1, 12.2) &-12.15 (-12.19, -12.11)\\ 
    \midrule
    Patient 6 &0.535 &0.684 &378730 &8.01(7.73, 8.26) &-12.13 (-12.16, -12.11)\\
    \midrule
    Patient 7 &0.633 &1.10 &241769 &11.8(11.7, 11.9) &-11.768(-11.772, -11.765)\\
    \midrule
    $^*$ & $\times 10^{-6}$ \\
    \end{tabular}
}
\end{table}

\begin{table}
  \centering
  {\small
  \caption{Inference results for clonality and entropy of memory T cells}\label{tab:naive}
\begin{tabular}{ccccll}
    \toprule
    &$\hat{a}$ &$\hat{b}$& $\hat{C}$ &Clonality(95\% CI)$^*$  &Entropy(95\% CI)\\

    \midrule
     Patient 1 &0.171 &0.301 &242994 &6.06(2.61, 8.73) &-10.79(-11.01, -10.52)\\
    \midrule
     Patient 2 &0.126 &0.132 &415556 &9.21(8.90, 9.50) &-10.84(-10.88, -10.79)\\ 
    \midrule
     Patient 3 &0.086 &0.111 &399632 &239(235, 244) &-9.66 (-9.67, -9.65)\\
    \midrule
     Patient 4 &0.113 &0.142 &443870 &37.0(35.3, 38.5) &-10.59(-10.65, -10.52)\\
    \midrule
     Patient 5 &0.134 &0.184 &391341 &9.70(9.10, 10.3) &-10.83(-10.88, -10.77)\\ 
    \midrule
    Patient 6 &0.150 &0.204 &439771 &7.14(6.74, 7.48) &-11.11(-11.16, -11.07)\\
    \midrule
    Patient 7 &0.163 &0.293 &204378 &33.5(31.6, 35.5) &-10.448(-10.452, -10.445)\\
    \midrule
    $^*$ & $\times 10^{-5}$ \\
    \end{tabular}
}
\end{table}

\section{Discussion}\label{sec:Discussion}
In this paper, we have discussed a method for constructing confidence intervals for entropy and clonality: both are functions of a high dimensional probability vector. The method was developed under a general Empirical Bayesian framework and thus is parametric in nature. As a consequence, the performance of the confidence interval depends on relevant parametric assumptions and the gamma distribution for the intensity rate in particular, even though the simulation study demonstrates certain robustness in the performance of constructed confidence intervals. One generalization is to replace this gamma distribution by a more flexible distribution.  For example, it is appealing to consider distributions from a nonparametric exponential family:
$p(\lambda \mid \eta)\propto p_0(\lambda)\exp\{B(\lambda)'\eta\},$ where $p(\cdot\mid \eta)$ is the density function of the intensity rate, and $B(\lambda)$ is a set of flexible basis functions given a priori such as $B(\lambda)=(\lambda, \lambda^2, \lambda^3)'$ \citep{schwartzman2008empirical}. The extension in this direction warrants further study. Lastly, in the proposed approach, the actual number of distinct clones was replaced by its estimator, which may affect the performance of the subsequent point and interval estimation. It is conceivable that the impact is greater for some functions such as entropy which is more sensitive to small-size clones than other functions such as clonality which is robust to small-size clones. However, estimating the number of distinct clones is analogous to estimating the number of unseen species, which is a difficult problem and in the current case depends on the parametric assumption for the intensity rate \citep{efron1976estimating}. Therefore, it is important to study the impact of this estimator on the construction of the confidence interval for different diversity parameters.

\bibliographystyle{imsart-nameyear} 
\bibliography{bibliography}       






\end{document}